\documentclass[aps,preprint,groupedaddress,showpacs,floatfix]{revtex4}
\usepackage{graphicx}
\begin{document}

\title{Transition form factors $\gamma^*\to \gamma f_2(1270)$ and $\gamma^*\to \gamma a_2(1320)$ in the $e^+e^-$ collisions}
\author {
N.N. Achasov$^{\,a}$ \email{achasov@math.nsc.ru} and A.V.
Kiselev$^{\,a,b}$, \email{kiselev@math.nsc.ru}}

\affiliation{
   $^a$Laboratory of Theoretical Physics,
 Sobolev Institute for Mathematics, 630090, Novosibirsk, Russia\\
$^b$Novosibirsk State University, 630090, Novosibirsk, Russia}

\date{\today}

\begin{abstract}
The predictions for transition form factors $\gamma^*\to \gamma
f_2(1270)$ and $\gamma^*\to \gamma a_2(1320)$ and corresponding
$e^+e^-\to \gamma^*\to f_2\gamma$ and $e^+e^-\to \gamma^*\to
a_2\gamma$ cross sections are obtained for the energy region up to
$2$ GeV. These predictions are coordinated with the recent Belle
data on the $\gamma^*(Q^2)\gamma\to f_2$ transition. It is shown
that the QCD asymptotics of the amplitudes of the reactions
$e^+e^-\to \gamma^*\to f_2\gamma$ and $e^+e^-\to \gamma^*\to
a_2\gamma$ can be reached only by taking into account a
compensation of contributions of $\rho(770)$, $\omega(782)$ with
contributions of their radial excitations. The relation
$\sigma(e^+e^-\to \gamma^*\to f_2\gamma)/\sigma(e^+e^-\to
\gamma^*\to a_2\gamma)\approx 25/9$, obtained with the help of the
$q\bar q$ model of $a_2$ and $f_2$ and QCD, is used to determine
the $\sigma(e^+e^-\to \gamma^*\to a_2\gamma)$ at high energies.
Recent BABAR measurement of the $e^+e^-\to
f_2\gamma\to\pi^+\pi^-\gamma$ cross section at $10.58$ GeV gives
hope for detailed investigation of the $\gamma^*(s)\to f_2\gamma$
and $\gamma^*(s)\to a_2\gamma$ transition form factors at high
energy region.

\end{abstract}
 \maketitle

\section{Introduction}

The $\gamma^*\to f_2(1270)\gamma$ and $\gamma^*\to
a_2(1320)\gamma$ transition form factors manifest themselves in
different reactions. In $e^+e^-\to\gamma^*\to \gamma f_2(a_2)$
processes they are in the time-like region of $\gamma^*$, while in
$e^+e^-\to e^+e^- \gamma^*\gamma \to e^+e^- f_2(a_2)$ reactions
these form factors are in the space-like region of $\gamma^*$.
These reactions are connected with each other, so progress in the
investigation of one reaction can shed light of the others.

Recently $\gamma^*(Q^2)\gamma\to f_2$ transition form factor in
the space-like region have been measured at Belle for $Q^2$ up to
$30$ GeV$^2$ \cite{BellePi0Pi0}. Besides, BABAR collaboration
measured the $e^+e^-\to f_2\gamma\to\pi^+\pi^-\gamma$ cross
section at $10.58$ GeV \cite{babar}, unfortunately, with large
error. The last result shows that it is possible to investigate
the $\gamma^*(s)\to f_2\gamma$ and $\gamma^*(s)\to a_2\gamma$
transition form factors in the time-like region at extremely high
energy ($s=-Q^2\approx 100$ GeV$^2$ and higher) in detail.

It will be shown below that these transition form factors are of
interest in the whole accessible $Q^2$ region. The point is that
the QCD asymptotics can be reached only by taking into account
cancellation of contributions of $\rho(770)$ and $\omega(782)$
with contributions of their radial excitations.

The Belle data on the $\gamma^*(Q^2)\gamma\to f_2$ transition form
factor was analyzed in Ref. \cite{aksh-2015}. In comparison to
that paper, we take into account $\rho''$ and $\omega''$
contribution in the $\gamma^*$ leg (the $\rho,\omega,\rho'$ and
$\omega'$ were taken into account in Ref. \cite{aksh-2015}). This
leads to modification of the Eq. (10) of the Ref. \cite{aksh-2015}
only.

The $e^+e^-\to\gamma^*\to \gamma f_2(a_2)$ cross sections were
estimated in Ref. \cite{agkr-2013} under assumptions
$g_{f_2\rho\rho}=g_{f_2\rho'\rho'}=g_{f_2\rho''\rho''}$ and
$g_{a_2\rho\omega}=g_{a_2\rho'\omega'}=g_{a_2\rho''\omega''}$. The
recent experimental data, the requirement of QCD asymptotics and
the naive quark model relations taken into account allow to make
prediction on quite a different level.

In Sec. \ref{Sec_eetof2g} the $e^+e^-\to\gamma^*\to f_2\gamma$
cross section is presented. It is shown that the cross section
should be larger than the Ref. \cite{agkr-2013} estimation, it
could be measured at the energies $\sim 2$ GeV at modern
colliders, for example, at VEPP-2000. Our estimation of the cross
section at the BABAR energy is smaller than the BABAR result.

In Sec. \ref{Sec_eetoa2g} we present the $e^+e^-\to\gamma^*\to
a_2\gamma$ cross section. In the $q\bar q$ model for $f_2$ and
$a_2$, $a_2=(u\bar u-d\bar d)/\sqrt{2}$ and $f_2=(u\bar u+d\bar
d)/\sqrt{2}$, the $\gamma^*(Q^2)\to f_2\gamma$ and
$\gamma^*(Q^2)\to a_2\gamma$ transition form factors are related
as $5:3$ at high $Q^2$ (in agreement with QCD), one could estimate
the $e^+e^-\to\gamma^*\to a_2\gamma$ cross section at high
energies using this relation.

In Sec. \ref{conclusion} the brief conclusion is presented and the
perspectives are discussed.

\section{The $e^+e^-\to f_2\gamma$ and $e^+e^-\to a_2\gamma$ cross sections}

\subsection{The $\sigma_{e^+e^-\to
f_2\gamma}(s)$}\label{Sec_eetof2g}

It is known that in the reaction $\gamma\gamma \to f_2\to \pi\pi$
tensor mesons are produced mainly by the photons with the opposite
helicity states. The effective Lagrangian in this case is

\begin{equation}
L=g_{f_2
\gamma\gamma}T_{\mu\nu}F_{\mu\sigma}F_{\nu\sigma}\,,\label{lagrfgg}
\end{equation}
$$F_{\mu\sigma}=\partial_\mu A_\sigma - \partial_\sigma A_\mu $$

\noindent where $A_{\mu}$ is a photon field and $T_{\mu\nu}$ is a
tensor $f_2$ field. For $\gamma^*(s)\to f_2\gamma$ transition
\begin{equation}g_{f_2 \gamma\gamma}\to g_{f_2 \gamma^*\gamma}(s)=F_{\gamma^*\to f_2\gamma
}(s)g_{f_2 \gamma\gamma}\end{equation}

\noindent where $F_{\gamma^*\to f_2\gamma }(s)$ is the transition
form factor.

In the frame of Generalized Vector Dominance Model (GVDM) we
assume that the effective Lagrangian of the reaction $f_2\to VV$
is \cite{akar-86}: \begin{equation} L=g_{f_2
VV}T_{\mu\nu}F^V_{\mu\sigma}F^V_{\nu\sigma}\,,\label{lagrfvv}
\end{equation}
$$F^V_{\mu\sigma}=\partial_\mu V_\sigma-\partial_\sigma V_\mu \,
.$$ \noindent where
$V=\rho,\,\rho',\rho''\,\omega,\,\omega',\omega''$. It is assumed
that $g_{f_2 \rho\rho'}$ and other constants with crossed vector
mesons are suppressed due to small overlap of the spatial wave
functions of $\rho,\rho',\rho''$ and similar for
$\omega,\omega',\omega''$:
\begin{equation}
g_{f_2 \rho\rho'}=0,\ g_{f_2
\omega\omega'}=0,..\label{crossZero}\end{equation}

Assuming the GVDM mechanism $e^+e^-\to
(\rho+\rho'+\rho''+\omega+\omega'+\omega'')\to f_2
(\rho+\rho'+\rho''+\omega+\omega'+\omega'')\to f_2 \gamma$ one
obtains \cite{agkr-2013} $$\sigma_{e^+e^-\to
f_2\gamma}(s)=\frac{4\pi^2}{9}\,\alpha^3
\Big(1-\frac{m_{f_2}^2}{s}\Big)^3\Big(\frac{s^2}{m_{f_2}^4}+3\frac{s}{m_{f_2}^2}+6\Big)\times
$$ \begin{equation} \bigg|\frac{m_\rho^2 g_{f_2\rho\rho}}{f^2_\rho
D_\rho(s)}+ \frac{m_{\rho'}^2 g_{f_2\rho'\rho'}}{f^2_{\rho'}
D_{\rho'}(s)}+ \frac{m_{\rho''}^2
g_{f_2\rho''\rho''}}{f^2_{\rho''} D_{\rho''}(s)}+\frac{m_\omega^2
g_{f_2\omega\omega}}{f^2_\omega D_\omega(s)}+ \frac{m_{\omega'}^2
g_{f_2\omega'\omega'}}{f^2_{\omega'} D_{\omega'}(s)}+
\frac{m_{\omega''}^2 g_{f_2\omega''\omega''}}{f^2_{\omega''}
D_{\omega''}(s)}\bigg|^2, \label{crossf2} \end{equation}

Here $D_V(s)$ are inverse propagators of vector mesons V:
\begin{equation} D_V(s)=m_V^2-s-i\sqrt{s}\Gamma_V(s),\label{Vprop}
\end{equation}
\noindent the forms of widths $\Gamma_V(s)$ are described below.
The constants $f_V$ are related to $e^+ e^-$ widths as usual:
\begin{equation}
\Gamma (V\to e^+ e^-)=\frac{4\pi\alpha^2}{3f_V^2}m_V.
\end{equation}

Let us denote $\rho=\rho_1, \rho'=\rho_2, \rho''=\rho_3$ and
similar for $\omega$, $\omega'$, $\omega''$. In the $q\bar q$
model for $i=1,2,3$
\begin{equation} g_{f_2\rho_i\rho_i}=g_{f_2\omega_i\omega_i},\,
m_{\omega_i}=m_{\rho_i},\,
f_{\omega_i}=3f_{\rho_i}.\label{rhoOmegaRelations}
\end{equation}

The $f_2\to\gamma\gamma$ width is \cite{agkr-2013}
\begin{equation} \Gamma_{f_2\to\gamma\gamma}
=\bigg(\frac{10}{9}\bigg)^2\,
\frac{\pi\alpha^2}{5}\bigg|\frac{g_{f_2\rho\rho}}{f^2_\rho}+
\frac{g_{f_2 \rho'\rho'}}{f^2_{\rho'}}+\frac{g_{f_2
\rho''\rho''}}{f^2_{\rho''}}\bigg|^2 m_{f_2}^3 = 3.03\pm 0.35\
\mbox{keV \cite{pdg-2014}},\label{f2gg}
\end{equation}

\noindent $\Gamma_{f_2\to\gamma\gamma}\equiv
\Gamma_{f_2\to\gamma\gamma}(m_{f_2}^2)$ and the factor $(10/9)^2$
takes into account $\omega_i$ contributions.

The QCD asymptotics of $\sigma(e^+e^-\to light\ hadrons)$ at
$s\to\infty$ is equal to $\sum Q_f^2 4\pi\alpha^2/s$ to within
logarithms, where $Q_f$ are charges of $u$, $d$, and $s$ quarks.
Due to asymptotic freedom at $s\to\infty$ imaginary part of
$D_V(s)$ vanishes
\begin{equation}
Im(D_V(s))=-\sqrt{s}\Gamma_V(s)\to 0\label{ImZero}\end{equation}

\noindent and

\begin{equation}\frac{1}{D_V(s)}\to \frac{1}{m_V^2-s}\,.\label{propAs}\end{equation} It means that

\begin{equation}\sigma_{e^+e^-\to f_2\gamma}(s)= A^2\bigg(\frac{10}{9}\bigg)^2\, \frac{4\pi^2}{9
m_{f_2}^4}\,\alpha^3+O(\frac{1}{s})\,,\end{equation}

\noindent so
\begin{equation}A=\frac{g_{f_2\rho\rho}m_\rho^2}{f^2_\rho}+
\frac{g_{f_2 \rho'\rho'}m_{\rho'}^2}{f^2_{\rho'}}+\frac{g_{f_2
\rho''\rho''}m_{\rho''}^2}{f^2_{\rho''}}\end{equation}

\noindent should be equal to zero, i.e. radial vector excitations
contribution compensates the contribution of vector bound state
$\rho$ in the leading order:
\begin{equation}
\frac{g_{f_2\rho\rho}m_\rho^2}{f^2_\rho}+ \frac{g_{f_2
\rho'\rho'}m_{\rho'}^2}{f^2_{\rho'}}+\frac{g_{f_2
\rho''\rho''}m_{\rho''}^2}{f^2_{\rho''}}=0. \label{zeroing}
\end{equation}

Note the analogous condition was obtained in Refs.
\cite{aks-2015,aksh-2015} after taking into account the QCD based
asymptotics of the $\gamma^*(Q^2)\gamma\to a_2$ amplitude.

Denoting like in Ref. \cite{aks-2015}
$a_{f_2}=g_{f_2\rho'\rho'}f_\rho^2/g_{f_2\rho\rho}f_{\rho'}^2$ and
$b_{f_2}=g_{f_2\rho''\rho''} f_\rho^2/g_{f_2\rho\rho}
f_{\rho''}^2$ the requirement Eq. (\ref{zeroing}) may be rewritten
as
\begin{equation}
m_\rho^2+a_{f_2}m_{\rho'}^2+b_{f_2}m_{\rho''}^2=0.
\label{zeroing2}
\end{equation}

In these terms Eqs. (\ref{crossf2}) and (\ref{f2gg}) give

$$\sigma_{e^+e^-\to f_2\gamma}(s)=\frac{4\pi^2}{9}\,\alpha^3
\Big(1-\frac{m_{f_2}^2}{s}\Big)^3\Big(\frac{s^2}{m_{f_2}^4}+3\frac{s}{m_{f_2}^2}+6\Big)\times
$$ $$\bigg|\frac{m_\rho^2 g_{f_2\rho\rho}}{f^2_\rho D_\rho(s)}+
\frac{m_{\rho'}^2 g_{f_2\rho'\rho'}}{f^2_{\rho'} D_{\rho'}(s)}+
\frac{m_{\rho''}^2 g_{f_2\rho''\rho''}}{f^2_{\rho''}
D_{\rho''}(s)}+\frac{1}{9}\bigg(\frac{m_\rho^2
g_{f_2\rho\rho}}{f^2_\rho D_\omega(s)}+ \frac{m_{\rho'}^2
g_{f_2\rho'\rho'}}{f^2_{\rho'} D_{\omega'}(s)}+ \frac{m_{\rho''}^2
g_{f_2\rho''\rho''}}{f^2_{\rho''}
D_{\omega''}(s)}\bigg)\bigg|^2=$$

\begin{equation}\frac{20\,\pi}{9}\,\alpha\,\frac{\Gamma_{f_2\to\gamma\gamma}}{m_{f_2}^3}\big|F_{\gamma^*\to
f_2\gamma }(s)\big|^2
\Big(1-\frac{m_{f_2}^2}{s}\Big)^3\Big(\frac{s^2}{m_{f_2}^4}+3\frac{s}{m_{f_2}^2}+6\Big).
\label{crossf2Fin}
\end{equation}

Since $\omega_i$ contributions are suppressed, we neglect the
difference between $\Gamma_{\omega_i}$ and $\Gamma_{\rho_i}$ in
$D_{\omega_i}(s)$ to simplify the formula for transition form
factor:
\begin{equation}
F_{\gamma^*\to f_2\gamma
}(s)=\frac{1}{1+a_{f_2}+b_{f_2}}\,\bigg(\frac{m_\rho^2}{D_\rho(s)}+
a_{f_2}\frac{m_{\rho'}^2}{D_{\rho'}(s)}+ b_{f_2}\frac{m_{\rho''}^2
}{D_{\rho''}(s)}\bigg), \label{f2TFF}
\end{equation}

The parameters of vector radial excitations ($m_V$, $f_V$ and
coupling constants) are not well-established. For the aims of this
work in region $\sqrt{s}<2$ GeV we use propagators in the form Eq.
(\ref{Vprop}) with constant widths with parameters taken from Ref.
\cite{pdg-2014}:
\begin{eqnarray} m_\rho=775\ \mbox{MeV},\
m_{\rho'}=1465\ \mbox{MeV},\ m_{\rho''}=1720\ \mbox{MeV},\nonumber
\\ \Gamma_{\rho}=149\ \mbox{MeV},\ \Gamma_{\rho'}=400\ \mbox{MeV},\
\Gamma_{\rho''}=250\ \mbox{MeV}.\label{massesV}\end{eqnarray}

Note one can use more complicated forms for $D_V(s)$ from Ref.
\cite{achkozh}.

To find the $(a_{f_2},b_{f_2})$ pair we need one more source of
information. Recently the Belle data on the
$\gamma^*(Q^2)\gamma\to f_2$ transition form factor
\cite{BellePi0Pi0} were analyzed in Ref. \cite{aksh-2015} without
taking $\rho''$ into account, i.e. $b_{f_2}=0$ in terms of this
work. To take $\rho''$ into account it is enough to modify Eq.
(10) of that paper: \begin{equation}F_T(Q) \to
F_T(Q)=\frac{1}{1+a_{f_2}+b_{f_2}}\bigg(\frac{1}{1+\frac{Q^2}{m_\rho^2}}+\frac{a_{f_2}}{1+\frac{Q^2}{m_{\rho'}^2}}+\frac{b_{f_2}}{1+\frac{Q^2}{m_{\rho''}^2}}\bigg).\end{equation}

This gives two minimums of the $\chi^2$ function, constructed for
the Belle data: $b_{f_2}=0.082\pm 0.025$ and $b_{f_2}=0.740\pm
0.011$. So then we will consider three Cases,\\[1pt]

\noindent Case 1: $(a_{f_2},b_{f_2})=(-0.393,0.082)$ - best
minimum of the $\chi^2$ function ($\chi^2/n.d.f.=29.3/28$);

\noindent Case 2: $(a_{f_2},b_{f_2})=(-1.3,0.74)$ - second minimum
of the $\chi^2$ function
($\chi^2/n.d.f.=49.1/28$);\begin{equation} \mbox{Case 3:
}(a_{f_2},b_{f_2})=(-0.28,0)\mbox{ as in in Ref. \cite{aksh-2015},
}\rho''\mbox{ is absent } (\chi^2/n.d.f.=39.1/29).\label{Cases}
\end{equation}

In Case 1 the $\rho''$ contribution is small, it is not far from
Case 3. The $\chi^2$ value in Case 2 is worse than in Cases 1 and
3, but since our consideration should be treated as a guide, this
minimum is also interesting for investigation.

\begin{figure}
\begin{center}
\begin{tabular}{ccc}
\includegraphics[width=8cm]{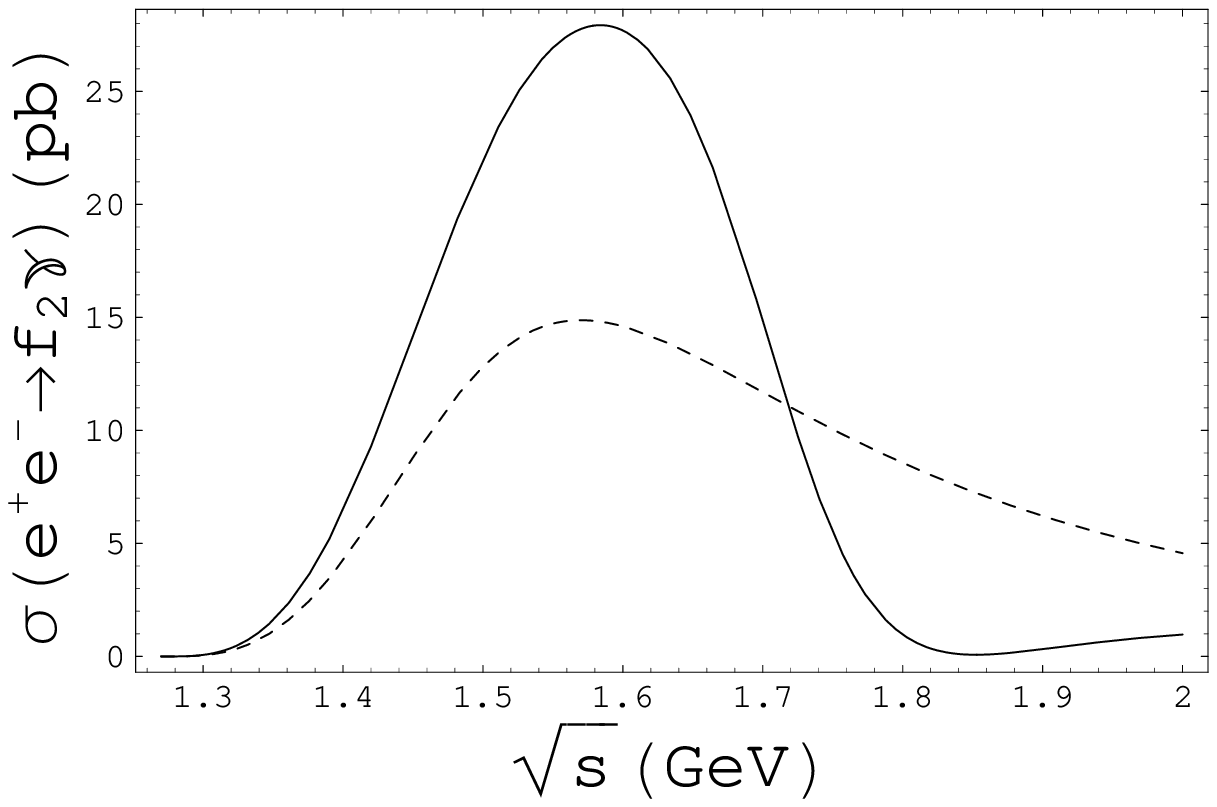} & \includegraphics[width=8cm]{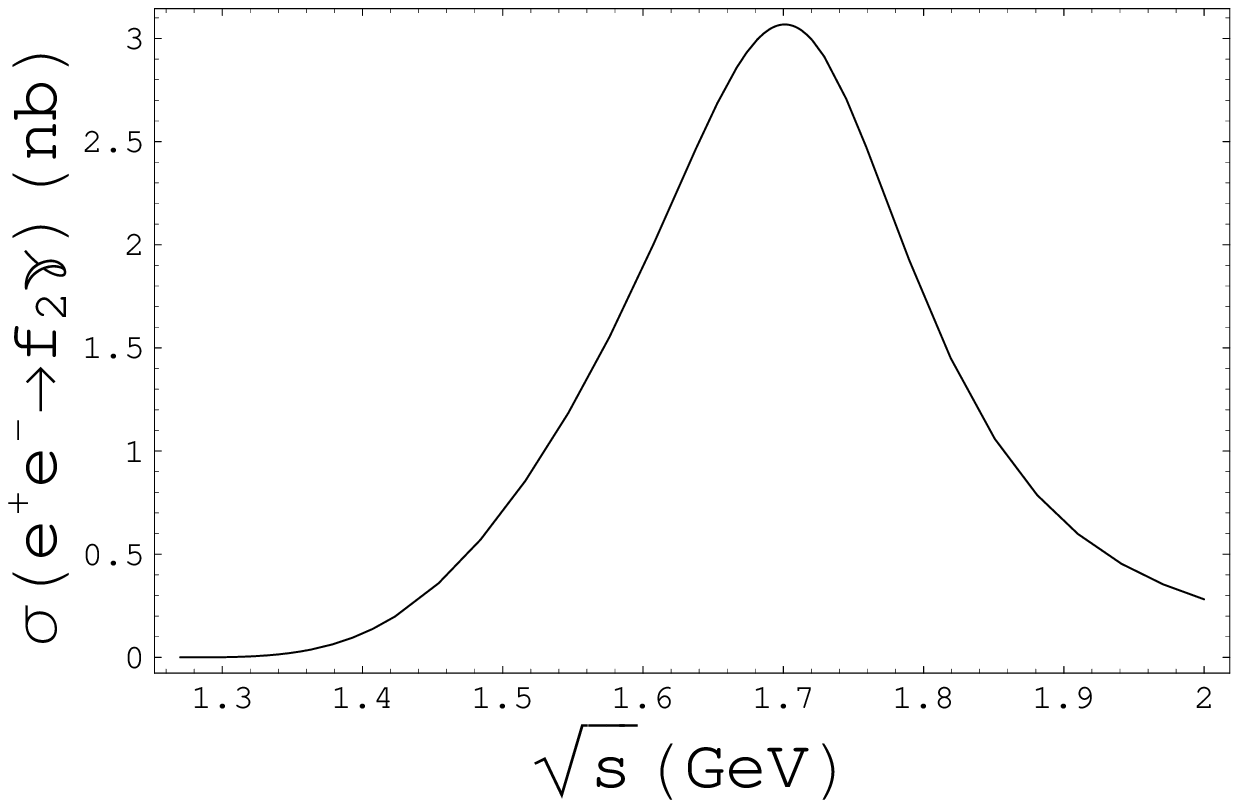}\\ (a)&(b)
\end{tabular}
\end{center}
\caption {The $e^+e^-\to f_2\gamma$ cross section. (a) Solid line
is for $(a_{f_2},b_{f_2})=(-0.393,0.082)$, dashed line is for
$(a_{f_2},b_{f_2})=(-0.28,0)$ ($\rho''$ is absent). (b) The plot
for $(a_{f_2},b_{f_2})=(-1.3,0.74)$.}
\label{eeF2Gamma}\end{figure}

All these three Cases are shown on Fig. \ref{eeF2Gamma}. In Case 3
(Fig. \ref{eeF2Gamma}b) the $\rho''$ peak is larger than in Fig.
\ref{eeF2Gamma}a at $\sim 100$ times.

The $e^+e^-\to f_2\gamma$ cross section is shown on Fig.
\ref{eeF2Gamma} for both minimums. One can see that Cases with
small and essential $\rho''$ contribution differ a lot: in Case 1
the cross section peak is $\sim 20$ pb, in the second one it is
two orders of magnitude higher. In both Cases this cross section
could be measured on the modern accelerators.

Note that maximum of the $e^+e^-\to f_2\gamma$ cross section was
estimated $5$ pb in Ref. \cite{agkr-2013}. The peak $\sim 3$ nb in
Case 2 is sensitive to the $\rho''$ width that is not well
established. If $\Gamma_{\rho''}$ was $500$ MeV (twice as much),
the peak would reduce to $0.8$ nb.

One can estimate relations between $f_2$ couplings to vector pairs
using values obtained in Ref. \cite{achkozh}
\begin{equation}
f_\rho=5.1, f_{\rho'}=5.0, f_{\rho''}=10.0. \label{fRho}
\end{equation}
\noindent and get
$$g_{f_2\rho\rho}:g_{f_2\rho'\rho'}:g_{f_2\rho''\rho''}=1:-0.41:0.33\mbox{\
in Case 1;} $$
$$g_{f_2\rho\rho}:g_{f_2\rho'\rho'}:g_{f_2\rho''\rho''}=1:-1.35:1.96\mbox{\
in Case 2;} $$
\begin{equation}
g_{f_2\rho\rho}:g_{f_2\rho'\rho'}:g_{f_2\rho''\rho''}=1:-0.29:0\mbox{\
in Case 3.}\label{gRelations1}
\end{equation}

Recently the $e^+e^-\to f_2\gamma\to\pi^+\pi^-\gamma$ cross
section was measured at BABAR at $10.58$ GeV:
$\sigma(\pi^+\pi^-\gamma)_{10.58}\equiv \sigma (e^+e^-\to
f_2\gamma\to\pi^+\pi^-\gamma,\ \mbox{10.58 GeV})=37^{+24}_{-18}$
fb \cite{babar}.

Taking into account $Br(f_2\to\pi^+\pi^-)\approx 56.5\%$ and Eq.
(\ref{propAs}), $D_V(s)\to m_V^2-s$ at $s\to\infty$, we find
$\sigma(\pi^+\pi^-\gamma)_{10.58}= 1.0$ fb for $b_{f_2}=0.082$ and
$\sigma(\pi^+\pi^-\gamma)_{10.58}= 3.8$ fb for $b_{f_2}=0.74$. The
elaboration of the pioneer measurement of
$\sigma(\pi^+\pi^-\gamma)_{10.58}$ is important for the
$\gamma^*\to\gamma f_2/a_2$ transition form factor investigations.

\subsection{The $\sigma_{e^+e^-\to a_2\gamma}(s)$}\label{Sec_eetoa2g}

The $\sigma_{e^+e^-\to a_2\gamma}$ treatment is similar
\cite{agkr-2013}:  $$\sigma_{e^+e^-\to
a_2\gamma}(s)=\frac{\pi^2}{9}\,\alpha^3
\Big(1-\frac{m_{a_2}^2}{s}\Big)^3\Big(\frac{s^2}{m_{a_2}^4}+3\frac{s}{m_{a_2}^2}+6\Big)\times
$$ \begin{equation}\bigg|\frac{m_\rho^2 g_{a_2\rho\omega}}{f_\rho
f_\omega D_\rho(s)}+ \frac{m_\omega^2 g_{a_2\rho\omega}}{f_\rho
f_\omega D_\omega(s)}+\frac{m_{\rho'}^2
g_{a_2{\rho'\omega'}}}{f_{\rho'} f_{\omega'} D_{\rho'}(s)}+
\frac{m_{\omega'}^2 g_{a_2\rho'\omega'}}{f_{\rho'} f_{\omega'}
D_{\omega'}(s)}+\frac{m_{\rho''}^2
g_{a_2{\rho''\omega''}}}{f_{\rho''} f_{\omega''} D_{\rho''}(s)}+
\frac{m_{\omega''}^2 g_{a_2\rho''\omega''}}{f_{\rho''}
f_{\omega''} D_{\omega''}(s)}\bigg|^2\label{crossa2}\end{equation}

\noindent and

\begin{equation} \Gamma_{a_2\to\gamma\gamma} =
\frac{\pi\alpha^2}{5}\bigg|\frac{g_{a_2\rho\omega}}{f_\rho
f_\omega}+\frac{g_{a_2\rho'\omega'}}{f_{\rho'}
f_{\omega'}}+\frac{g_{a_2\rho''\omega''}}{f_{\rho''}
f_{\omega''}}\bigg|^2 m_{a_2}^3 = 1.00\pm 0.06\ \mbox{keV
\cite{pdg-2014}}.\label{Ga2ToGG}\end{equation}

The QCD asymptotics requires

\begin{equation}
\frac{g_{a_2\rho\omega}}{f_{\rho} f_{\omega}}
m_\rho^2+\frac{g_{a_2\rho'\omega'}}{f_{\rho'}
f_{\omega'}}m_{\rho'}^2+\frac{g_{a_2\rho''\omega''}}{f_{\rho''}
f_{\omega''}}m_{\rho''}^2=0\,,\label{a2As}
\end{equation}
\noindent here Eqs. (\ref{rhoOmegaRelations}) were used. Denoting
$a_{a_2}=g_{a_2\rho'\omega'}f_\rho
f_\omega/g_{a_2\rho\omega}f_{\rho'} f_{\omega'}$ and
$b_{a_2}=g_{a_2\rho''\omega''}f_\rho
f_\omega/g_{a_2\rho\omega}f_{\rho''} f_{\omega''}$ the condition
Eq. (\ref{a2As}) may be rewritten as
\begin{equation}
m_\rho^2+a_{a_2}m_{\rho'}^2 +b_{a_2}m_{\rho''}^2=0.
\label{zeroing3}
\end{equation}

Note that in the naive $q\bar q$ model for $i = 1,2,3$
\begin{equation} g_{a_2\rho_i\omega_i}=2 g_{f_2\rho_i\rho_i}.\label{a2f2Consts}\end{equation}

Together with Eqs. (\ref{rhoOmegaRelations}) it means that
\begin{equation}a_{a_2}=a_{f_2},\,b_{a_2}=b_{f_2}.\label{abEqual} \end{equation}

Eqs. (\ref{rhoOmegaRelations}) lead to
\begin{equation}\sigma_{e^+e^-\to
a_2\gamma}(s)=\frac{20\pi}{9}\,\alpha\,\frac{\Gamma_{a_2\to\gamma\gamma}}{m_{a_2}^3}\Big|F_{\gamma^*\to
a_2\gamma}(s)\Big|^2\,
\Big(1-\frac{m_{a_2}^2}{s}\Big)^3\Big(\frac{s^2}{m_{a_2}^4}+3\frac{s}{m_{a_2}^2}+6\Big).\label{crossa2Final}\end{equation}

\noindent with transition form factor
\begin{equation}F_{\gamma^*\to a_2\gamma}(s)=\frac{1}{2(1+a_{a_2}+b_{a_2})}\bigg(\frac{m_\rho^2}{D_\rho(s)}+
\frac{m_\rho^2}{D_\omega(s)}+a_{a_2}\frac{m_{\rho'}^2}{D_{\rho'}(s)}+
a_{a_2}\frac{m_{\rho'}^2}{D_{\omega'}(s)}+b_{a_2}\frac{m_{\rho''}^2
}{D_{\rho''}(s)}+
b_{a_2}\frac{m_{\rho''}^2}{D_{\omega''}(s)}\bigg)\label{a2TFF}\end{equation}

QCD also reqiures at high $s$
\begin{equation} \frac{\sigma (e^+e^-\to f_2\gamma,s)}{\sigma
(e^+e^-\to a_2\gamma,s)}=\frac{25}{9}.\label{dev25}
\end{equation}

It's easy to see that our $q\bar q$ consideration, Eqs.
(\ref{rhoOmegaRelations}), (\ref{crossZero}), (\ref{propAs}) and
(\ref{a2f2Consts}), agrees with Eq. (\ref{dev25}).

It is natural that our consideration also provides the relation

\begin{equation}
\frac{\Gamma(f_2\to\gamma\gamma)}{\Gamma(a_2\to\gamma\gamma)}=\frac{25}{9}\frac{m_{f_2}^3}{m_{a_2}^3}\approx\frac{25}{9}\,,\label{rel_a2_b2}
\end{equation}

\noindent well-known in the $q\bar q$ model, which is in perfect
agreement with the data, see Eqs. (\ref{f2gg}) and
(\ref{Ga2ToGG}).

Note that using the above-mentioned BABAR measurement $\sigma
(e^+e^-\to f_2\gamma\to\pi^+\pi^-\gamma,\ \mbox{10.58
GeV})=37^{+24}_{-18}$ fb, $Br(f_2\to\pi^+\pi^-)\approx 56.5\%$ and
Eq. (\ref{dev25}) one can obtain
\begin{equation}
\sigma (e^+e^-\to a_2\gamma,\mbox{ 10.58
GeV})=24^{+15}_{-11}\mbox{ fb}. \label{babar_a2}
\end{equation}

\begin{figure}
\begin{center}
\begin{tabular}{ccc}
\includegraphics[width=8cm]{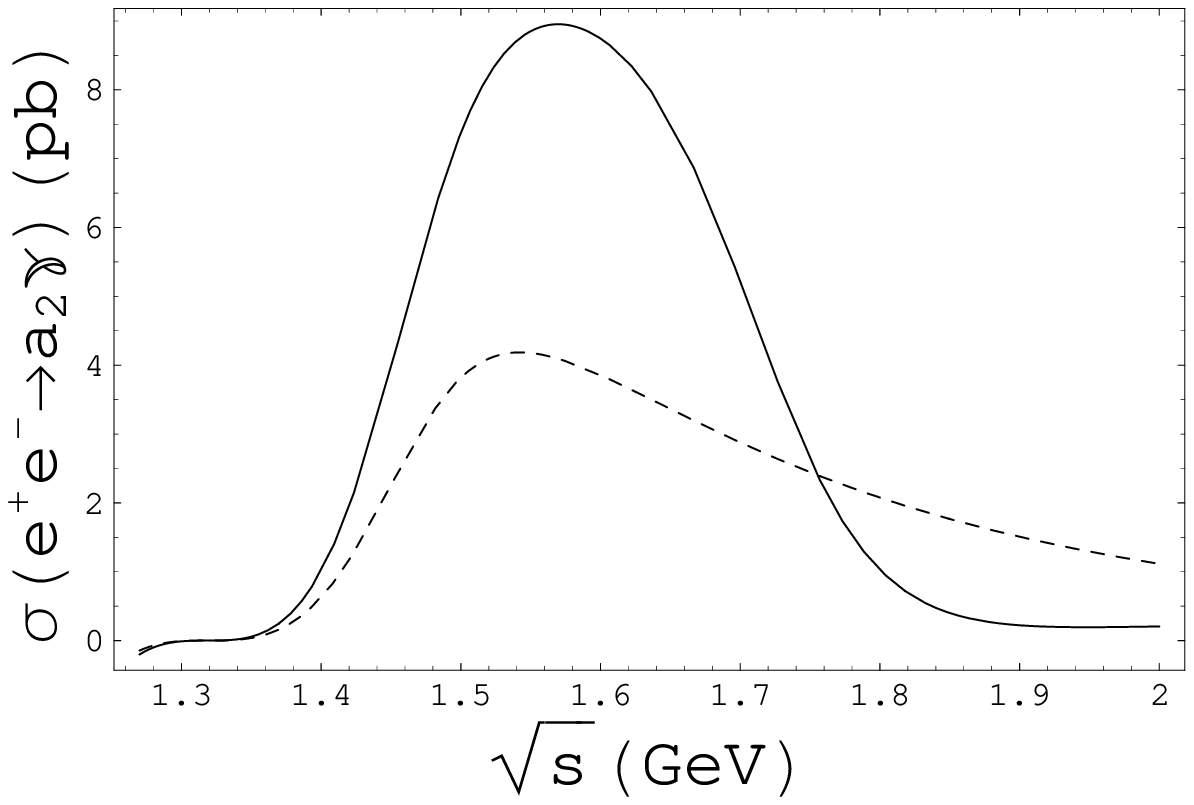} & \includegraphics[width=8cm]{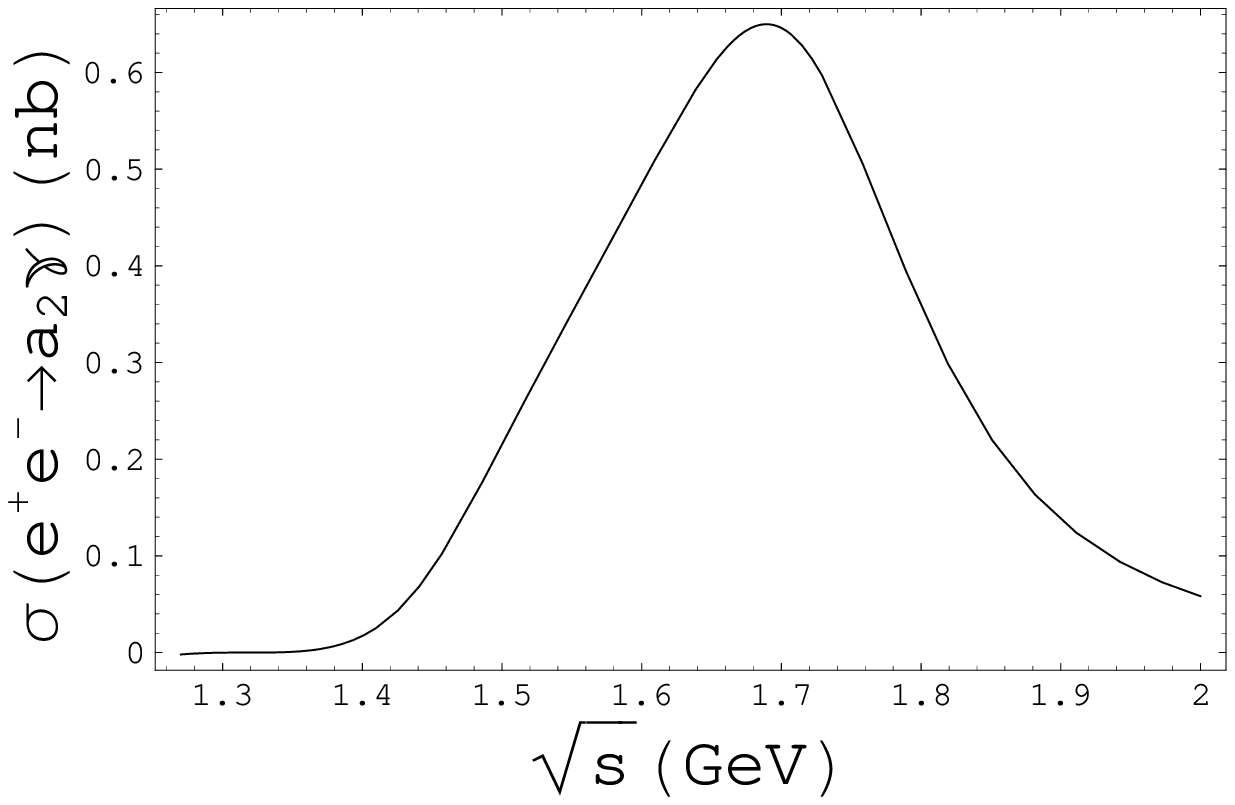}\\ (a)&(b)
\end{tabular}
\end{center}
\caption {The $e^+e^-\to a_2\gamma$ cross section. (a) Solid line
is for $(a_{a_2},b_{a_2})=(-0.393,0.082)$, dashed line is for
$(a_{a_2},b_{a_2})=(-0.28,0)$. (b) The plot for
$(a_{a_2},b_{a_2})=(-1.3,0.74)$.}
 \label{eeA2Gamma}\end{figure}

On the Fig. \ref{eeA2Gamma} we show the $\sigma (e^+e^-\to
a_2\gamma,s)$ with sets $(a_{a_2},b_{a_2})$ as in Eqs.
(\ref{Cases}), see Eq. (\ref{abEqual}). We take \cite{pdg-2014}

\begin{equation}
\Gamma_{\omega}=8.5\ \mbox{MeV},\ \Gamma_{\omega'}=215\
\mbox{MeV},\ \Gamma_{\omega''}=315\
\mbox{MeV}.\label{GammasOmegas}\end{equation}

In our approach
$g_{a_2\rho\omega}:g_{a_2\rho'\omega'}:g_{a_2\rho''\omega''}$
relations are the same as Eq. (\ref{gRelations1}).

\section{Conclusion}\label{conclusion}

It is shown that the QCD asymptotics of the amplitudes of the
reactions $e^+e^-\to \gamma^*\to f_2\gamma$ and $e^+e^-\to
\gamma^*\to a_2\gamma$ can be reached only by taking into account
a compensation of contributions of $\rho(770)$, $\omega(782)$ with
contributions of their radial excitations. It is shown also that
the $\rho$ and $\omega$ excitation contribution to the $e^+e^-\to
f_2\gamma$ and $e^+e^-\to a_2\gamma$ cross sections is essential
and allows the experimental investigation of
$e^+e^-\to\eta\pi^0\gamma$ and $e^+e^-\to\pi^0\pi^0\gamma$
processes, for example, at VEPP-2000 collider. Note that the best
channel to study the $a_2$ production is the $e^+e^-\to
a_2\gamma\to\rho\pi\gamma$ because $Br(a_2\to\rho\pi)=70\%$
\cite{pdg-2014} and the background is expected to be small.

At the BABAR energy $10.58$ GeV our estimation of the $e^+e^-\to
f_2\gamma\to\pi^+\pi^-\gamma$ cross section is much less than the
average value presented in \cite{babar}, but the experimental
error is half of the average value, so no conclusions concerning
this disagreement could be done now. Anyway, the pioneer BABAR
measurement gives hope that the detailed investigation of the
$\gamma^*(s)\to f_2\gamma$ and $\gamma^*(s)\to a_2\gamma$
transition form factors at high energy region is possible.

Emphasize that the results obtained in Sec. II do not represent a
precise prediction, they should be treated as a guide.

\section{Acknowledgements}

This work was supported in part by RFBR, Grant No. 16-02-00065,
and the Presidium of RAS project No. 0314-2015-0011.

\end{document}